\documentclass[iop]{emulateapj}

\newcommand{\myemail}{\url{msabdeen@uark.edu}}
\newcommand {\apgt} {\ {\raise-.5ex\hbox{$\buildrel>\over\sim$}}\ }
\newcommand {\aplt} {\ {\raise-.5ex\hbox{$\buildrel<\over\sim$}}\ }

\shorttitle{Determining the Co-Rotation Radii of Spiral Galaxies Using Pitch Angle Measurements}

\shortauthors{Abdeen et al.}

\usepackage{color}
\definecolor{gray}{gray}{0.5}
\usepackage{url}
\usepackage[latin1]{inputenc}
\usepackage{tikz}
\usetikzlibrary{shapes,arrows}
\usepackage{epsfig}
\usepackage{amsmath}
\usepackage{hyperref}

\submitted{First Draft, Jan 30, 2020}

\begin{document}

\title{Determining the Co-Rotation Radii of Spiral Galaxies Using Spiral Arm Pitch Angle Measurements at Multiple Wavelengths}

\author{Shameer Abdeen \altaffilmark{1,2}, 
Daniel Kennefick\altaffilmark{1,2},
Julia Kennefick\altaffilmark{1,2},
Ryan Miller\altaffilmark{3},
Douglas W. Shields\altaffilmark{1,2},
Erik B. Monson\altaffilmark{1,2} and
Benjamin L. Davis\altaffilmark{4}}

\altaffiltext{1}{Department of Physics, University of Arkansas, 226 Physics Building, 825 West Dickson Street, Fayetteville, AR 72701, USA}
\altaffiltext{2}{Arkansas Center for Space and Planetary Sciences, University of Arkansas, 346 1/2 North Arkansas Avenue, Fayetteville, AR 72701, USA; \myemail}
\altaffiltext{3}{Department of Physics, Utica College, 1600 Burrstone Rd, Utica, NY 13502, USA;}
\altaffiltext{4}{Centre for Astrophysics and Supercomputing, Swinburne University of Technology, Hawthorn, VIC 3122, Australia}

\begin{abstract}
The spiral arms spanning disk galaxies are believed to be created by density waves
that propagate through galactic disks. We present a novel method of finding the co-rotation
radius where the spiral arm pattern speed matches the velocities of the stars within the disk. Our
method uses an image-overlay technique, which involves tracing the arms of spiral galaxies on
images observed in different wavelengths. Density wave theory predicts that spiral arms observed
from different wavelengths show a phase crossing at the co-rotation radius. For the purpose of this
study, 20 nearby galaxies were analyzed in four different wavelengths with pitch angle measurements
performed by two independent methods. We used optical wavelength images ($B$-band 440nm), two
infrared wavelength images provided by Spitzer (3.6$\mu$m and 8$\mu$m) and ultraviolet images
from GALEX (1350\AA, 1750\AA). The results were compared and verified with other records found in the
literature. We then found rotation curve data for six of our galaxies and used our co-rotation radii
estimates to measure the time that would elapse between star formation and moving to their observed
positions in the $B$-band spirals. The average time lapse for this motion was found to be $\sim$ 50 Myr. The success of this new method of finding the co-rotation radius confirms density wave theory in a very direct way.
	
\end{abstract}



\section{Introduction} The density wave theory, introduced by C.C. Lin \& Frank Shu \citep {Lin:Shu:1964,Bertin:Lin:1996,Shu:2016} proposed a theoretical explanation for the existence of spiral structures in galaxies due to long lived quasi-stationary density waves. Subsequent development of the theory has focused on the highly dissipative nature of the
galactic disk, which casts doubt on the semi-permanence of the spiral patterns envisaged by the original theory \citep{Toomre:1969}. 
Numerical simulations of density waves in disks have especially fostered the view that spiral patterns created by 
density waves may be transient in nature \citep{Sellwood:Carlberg:1984}, possibly lasting only as long as one or two rotations of the galactic
disk (recall that the Milky Way has a rotation period of 250 Myr). Nevertheless some simulations continue to suggest
that spiral patterns could be quite long-lived, lasting many rotational periods \citep{Sellwood:2012, Sellwood:Carlberg:2014} . 

It is not only theorists who disagree on this topic. Observational studies have also contradicted each other upon
this point. \cite{Foyle:2011} point out that the absence of color gradients in the spiral arms of galaxies would
suggest that the density waves are too transient to permit time for such gradients to develop. But the work
of \cite{MartinezGarcia:GonzalezLopezlira:2015} does find evidence for such color gradients and they argue
that the spiral arms must be long-lived. Evidence for these color gradients is presented in \cite{Pour-Imani:2016,Rayan:2019} 
using a different approach, which demonstrates that the spiral arm pitch angle in wavebands associated
with star-formation is consistently larger than those associated with the stellar population. This suggests that
new-born stars are found downstream of the star-formation region in a manner consistent with the density-wave
theory. A study by \cite{Yu_2018} with a large number of galaxies strongly confirms this picture and agrees with the work of \cite{MartinezGarcia:GonzalezLopezlira:2015}. \cite{Pour-Imani:2016} also agrees with theirs in broad outline (but see Section 4.2 below for one significant detail where they differ). 

In this paper, we seek to take advantage of this discovery to further explore the nature of the spiral
structure. We show how overlaying the different spiral arms, from the star-forming region and from the stellar arm enables us to find the co-rotation radius (R$_{CR}$) of a galaxy. Then, for those galaxies for which 
rotation curves are available, we calculate the time that elapses between stars being born in the star-forming
region and then being observed downstream in the $B$-band. This time is found to be a significant fraction of the rotational period of a galaxy, thus suggesting, in agreement with \cite{MartinezGarcia:GonzalezLopezlira:2015}, that the spiral arms cannot be extremely transient.  

 Due to the differential rotation of the disk, stellar populations move faster than the spiral arm density wave inside the co-rotation radius, while they lag behind the spiral arm density wave outside the co-rotation radius. The general motion of stellar populations about the galactic center can be considered as a motion along an axially symmetric gravitational field, and with a proper selection of an appropriate reference frame, one can express this as an epicyclical motion. The inner Lindblad (ILR) and the outer Lindblad (OLR) resonances are defined in terms of this epicyclic frequency. The pattern speed  ($\Omega$$_{p}$) and the angular velocity of the disk $\Omega$(R) are related to the epicyclic frequency $\kappa$ by the equation, $\Omega$$_{p}$= $\Omega$(R) $\pm$ $\kappa$/2, where the plus and minus signs give the ILR and the OLR, respectively. 

Identifying the location of the co-rotation radius is of central importance to understanding the evolution of the spiral
structure. The existence of a co-rotation radius itself serves as evidence of the validity of the density wave theory. Furthermore, since the co-rotation radius acts as a region of resonance or a stable region within the galaxy structure, it has a dynamical influence on the matter and the chemical composition distribution process of the galaxy \citep{Scarano:Lepine:2013}. The region of stability around the co-rotation radius creates an ideal environment for a Galactic habitable zone (GHZ) e.g. \cite{Marochnik:1983, Sundin:2006}.There are multiple papers  \citep{Creze:Mennessie:1973,Mishurov:Zenina:1999} among others, that claim to find the location of the Sun close to the galactic co-rotation radius of the Milky Way. 

There are many methods of identifying the co-rotation radius. The Tremaine-Weinberg method \citep{Tremaine:Weinberg:1984} hereafter TW, requires certain specific morphological features along with the assumptions (a) the disc of the galaxy is flat, (b) the disc has a well-defined pattern speed, and (c) the surface brightness of the tracer obeys the continuity equation. The TW method directly measures the bar pattern speeds. Using the TW method some authors \citep{Merrifield:2006,Grand:2012} have claimed to observe multiple spiral arm pattern speeds that vary with the radius from the galactic center, which may result in observing multiple co-rotation radii. 

Finding the co-rotation radius from morphological features is also a common technique. \cite{Elmegreen:Elmegreen:1990} claimed to observe sharp end points to star formation ridges and dust lanes near the co-rotation radius in gas-rich galaxies. \cite{Patsis:2003} claimed that rings in spiral galaxies can be related to resonance locations.   

\cite{Scarano:Lepine:2013} analyzed the relationship between the galactic radius where the metallicity distributions break and the co-rotation radius.  Type II supernovae explosions associated with spiral arm density wave propagation are directly responsible for oxygen production. By analyzing oxygen abundance from line fluxes, they were able to find a convincing relationship between the co-rotation radius and the metallicity distribution break radius. 

The Puerari \& Dottori method \citep{Puerari:Dottori:1997} hereafter PD, uses a photometric method to find the phase crossings in Fourier transforms of azimuthal profiles in $B$ and $I$ bands. The motion of the spiral arm density wave through the galactic disk generates a shock induced star formation process that produces an azimuthal age gradient of stars throughout the spiral arm. \cite{Martinez-Garcia:2009} claimed to observe this age gradient in the form of a color gradient flowing azimuthally outwards across the spiral arm flowing in opposite directions on either side of the R$_{CR}$. Throughout history $B$, $I$, $HI$, $CO$ and $FUV$ wave bands have been used in order to observe this phase shift.  

In agreement with the PD method, we were able to introduce a novel way of determining the co-rotation radius, based on an image overlaying technique, which involves tracing the arms of spiral galaxies on images observed from different wavelengths. In a previous study \citep{Pour-Imani:2016}, convincing evidence was given in favor of an implication of density wave theory, which states that the pitch angle varies with wavelength for a given galaxy. Thus we have
different apparent spiral arms at different wavelengths. Since these spiral arms have different pitch angles, it is possible for them to cross over each other when they are overlaid on the same image of the galactic disk. The crossing-point should
be the co-rotation radius. Below we measure the co-rotation radius by this method and compare it to other methods
which have been used for our sample and find it is successful at correctly measuring the position of this radius.



\begin{deluxetable*}{llccllllc}
\tablecolumns{7}
\tablecaption{Sample\label{Sample}}
\tablehead{
\colhead{Galaxy Name } & \colhead{ Hubble Type }  & \colhead{ RA (J2000) } & \colhead{Dec(J2000)}& \colhead{Distance (Mpc)} & \colhead{ Method }& \colhead{Image Sources }  \\
\colhead{(1)} & \colhead{(2)} & \colhead{(3)} & \colhead{(4)} & \colhead{(5)}& \colhead{(6)}& \colhead{(7)}}

\startdata

	NGC 0628 & SA(s)c & 01 36 41.7 &+15 47 01 & 6.7 & SNII*&[1],[2],[3],[5]  \\ 
	NGC 2403 & SAB(s)cd &07 36 51.4 &+65 36 09 & 3.47 & SNII* &[1],[2],[3],[6]\\ 
	NGC 3031 & SA(s)ab&09 55 33.2 &+69 03 55 & 3.63 & Cph*&[1],[2],[3],[7] \\ 
	NGC 2903 & SAB(rs)bc& 09 32 10.1 &+21 30 03  & 6.52 & Hub*&[1],[2],[3],[8] \\ 
	NGC 4254 & SA(s)c &12 18 49.6 &+14 24 59 & 32.30 & Hub*&[1],[2],[4],[9]\\ 
	NGC 5194 & SA(s)bc&13 29 52.7 &+47 11 43 &8.9 & SNII*&[1],[2],[3],[8] \\ 
	NGC 5236 & SAB(s)c&13 37 00.9 &$-$29 51 56 &4.5 & Cph* &[1],[2],[3],[10]\\ 
	NGC 5457 & SAB(rs)cd &14 03 12.5 &+54 20 56 &7.4 & Cph* &[1],[2],[3],[8]\\ 
	NGC 1566 & SAB(s)bc&04 20 00.4 &$-$54 56 16  &18.2 & Hub* &[1],[2],[3],[11]\\ 
	NGC 4321 &  SAB(s)bc&12 22 54.8 & +15 49 19 &16.10 & Cph*&[1],[2],[3],[7] \\ 
	NGC 5033 & SA(s)c &13 13 27.4 &+36 35 38 &12.7 & Hub* &[1],[2],[8]\\ 
	NGC 6946 & SAB(rs)cd &20 34 52.3 &+60 09 14 & 3.76 & Hub* &[1],[2],[12]\\ 
	NGC 1042 & SAB(rs)cd &02 40 24.0 &$-$08 26 01 & 15.0 & Tully-Fisher &[1],[3],[12]\\ 
	NGC 4579 & SAB(rs)b &12 37 43.5 &+11 49 05 & 16.7 & Tully-Fisher &[1],[3],[12]\\ 
	NGC 5701 &(R)SB(rs)0/a &14 39 11.1 &+05 21 49 & 15.4 & Tully-Fisher &[1],[3],[13]\\ 
	NGC 5850 & SB(r)b &15 07 07.7  &+01 32 39  & 18.7 &  Tully-Fisher & [1],[2],[12]\\ 
	NGC 3938 & SA(s)c &11 52 49.4  &+44 07 15 & 21.9 & SNII & [1],[2],[8]\\ 
	NGC 4136 & SAB(r)c &12 09 17.7 &+29 55 39 & 9.7 & Tully-Fisher &[1],[3],[12]\\ 
	NGC 7479 & SB(s)c & 23 04 56.6  &+12 19 22 & 36.8 & Tully-Fisher & [1],[3],[12]\\ 
	NGC 7552 & (R')SB(s)ab & 23 16 10.7  & $-$42 35 05 & 14.8 & Tully-Fisher & [1],[2],[10]\\ 

\enddata

\tablecomments{Columns:
(1) Galaxy name
(2) Hubble morphological type
(3) RA (J2000)
(4) DEC (J2000)
(5) Distance (Mpc)
(6) Method for distance calculations 
(* Distance and  Method for distance calculations are based on the compilations of Scarano \& L{\'e}pine \citep{Scarano:Lepine:2013})
(7) Image Sources (Telescope/wavelength): 
[1]=IRAC 3.6$\mu$m,[2]=IRAC 8.0$\mu$m,[3]=GALEX 1516\AA,[4]=GALEX 1542\AA,
[5]=NOT 4400\AA,[6]=LCO 4400\AA,[7]=WIYN 4331\AA,[8]=KPNO 4400\AA,[9]=INT 4034\AA,
[10]=CTIO 4400\AA,[11]= duPont 4400\AA,[12]=JKT 4034\AA,[13] = Lowell1.1m 4500\AA
}
\end{deluxetable*}

\section{The Data Sample}
The accuracy of our method relies on three cardinal criteria. (a) The inclination of the galaxy. If the galaxy has a very high inclination angle, most of the crucial details will be lost in the de-projection process. We selected our galaxies such that the inclination angles are $\leq$ 60$^{\circ}$. (b) Well defined spiral structure. The overlaying technique requires the galaxy sample to have a well-defined, clear spiral structure in order to accurately overlay the synthetic spirals. (c) The variation of the spiral arm pitch angle with the galactic radius remains constant for a significant proportion of the galactic radius. This amounts to a criterion that the spiral arms are logarithmic spirals. Although we assign a mean value for the pitch angle, the extent to which the spiral arms are intrinsically logarithmic plays a vital role in this criterion.

Our galaxy sample (see Table 1) was selected based on the availability of co-rotation radii measurements in the literature, along with considering the above mentioned criteria. Twenty nearby galaxies were analyzed in four different wave bands. Twelve galaxies were selected from the galaxy sample of  Scarano \& L{\'e}pine \citep{Scarano:Lepine:2013} while the remaining eight galaxies were selected from \cite{Buta:Zhang:2009}. We used optical wavelength images ($B$-band 440nm), two infrared wavelengths (3.6$\mu$m and 8$\mu$m) from the Spitzer Infrared Nearby Galaxies Survey (SINGS) \footnote[1]{https://irsa.ipac.caltech.edu/data/SPITZER/SINGS/}, and ultraviolet images from Galaxy Evolution Explorer (GALEX) \footnote[2]{https://archive.stsci.edu/missions-and-data/galex-1/}. For 12 galaxies, the distances were  recorded based on the compilations of Scarano \& L{\'e}pine \citep{Scarano:Lepine:2013}. Our galaxy sample was limited to a distance of 36.8 Mpc (z = 0.007942), where NGC 7479 had the furthest recorded distance.


\section{Analysis}
\subsection{De-projecting and Measuring Pitch Angles }

Galaxies with non-zero inclination angles between the plane of sight and their disks had to be de-projected to a face-on orientation.  Since we were working with four different wavelengths, we first created a composite image. Using the composite image we used the traditional approach of fitting elliptical isophotes and then transformed them to a circular configuration. We used the de-projecting parameters of the composite image, namely the position angle and the ellipticity and applied them to each individual wavelength image. 

\begin{figure}
\label{fig1}
\includegraphics[width=8.6cm]{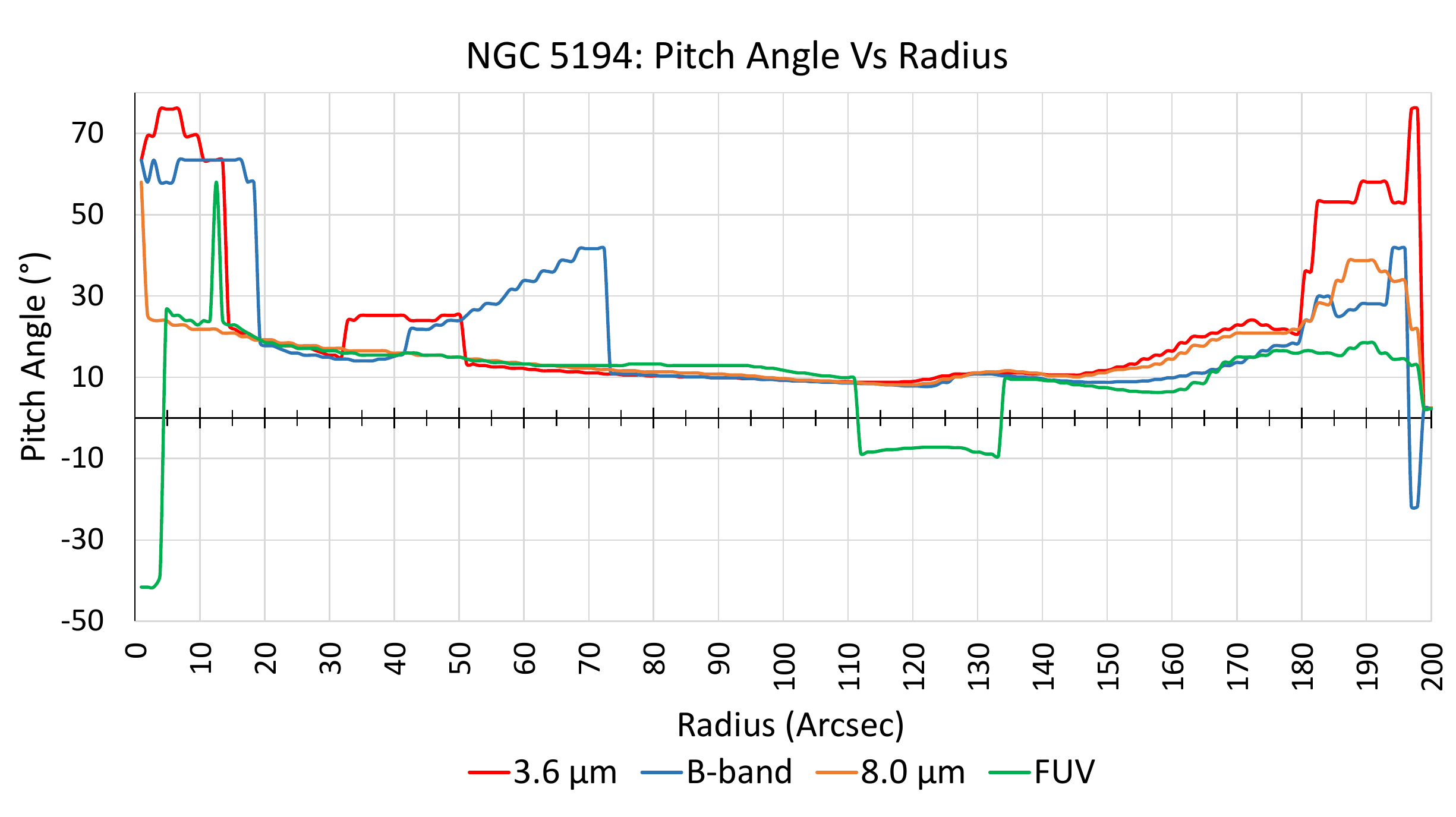}
\caption{The graph of the pitch angle variation, as a function of radius for NGC 5194.}

\includegraphics[width=8.6cm]{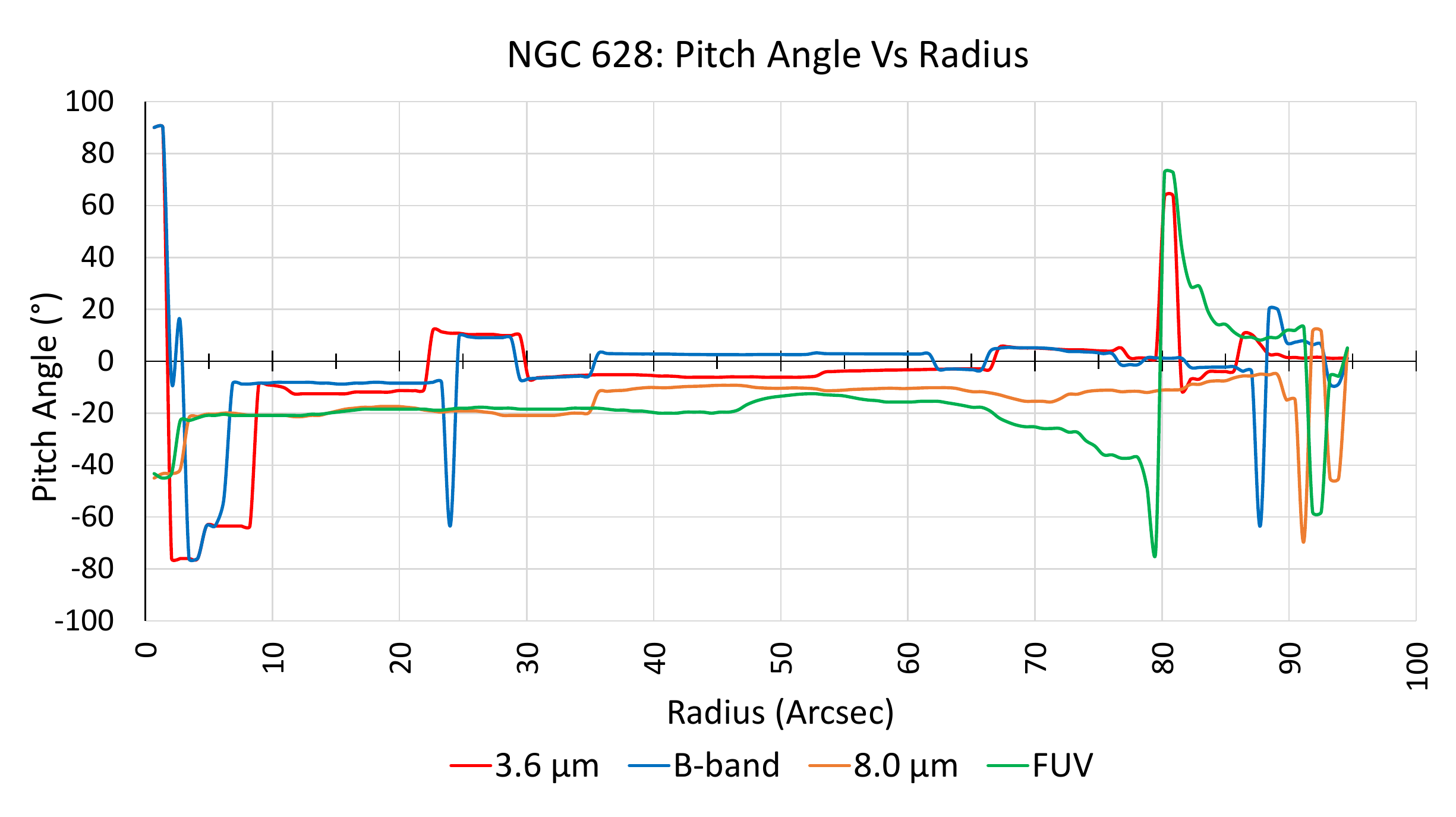}
\caption{The graph of the pitch angle variation, as a function of galaxy radius for NGC 628. The graphs in figure 1 and 2 were created using the results from the 2DFFT method.}
\label{fig2}
\end{figure}

\begin{figure}
	\includegraphics[width=8.6cm]{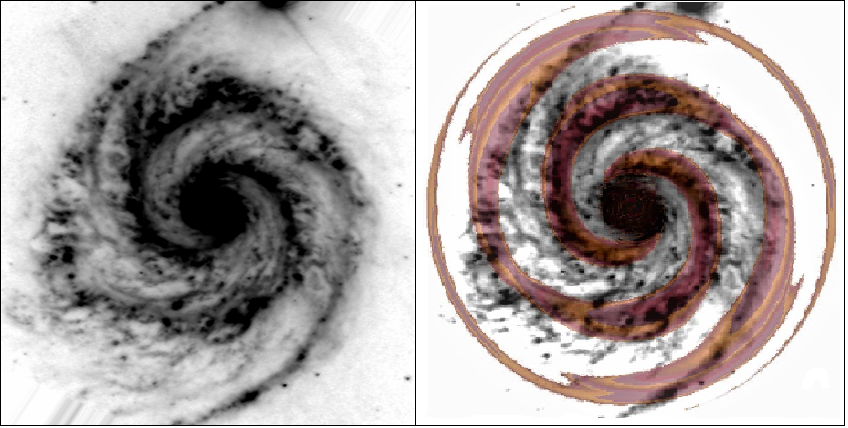}
	\caption{ NGC 5194 along with its inverse 2DFFT output overlaid on the original image. }
	\label{fig3}
\end{figure}

The pitch angle of a spiral is defined as the angle between the spiral\textquoteright s tangent at a given point and the tangent line drawn to a concentric circle  with the center located at the galactic center, that passes through that point. Pitch angle measurements were performed by two independent methods namely 2DFFT method \citep{Seigar:2008, Davis:2012} and the \textsc{Spirality} \citep{Shield:2015} \footnote[3]{Spirality: http://ascl.net/phpBB3/download/file.php?id=29} method. The 2DFFT method uses a modified two dimensional fast Fourier transform process to generate the pitch angles as a function of galaxy radius in pixels for different harmonic modes. Figure 1 shows the variation of pitch angle with radius (galactocentric offsets) in NGC 5194 for the second harmonic mode. NGC 5194 is an example of a galaxy that has a small relative difference in the variation of the pitch angles in all four wavelengths, hence giving rise to nearly equal mean pitch angles (see NGC 5194 in Table2) while NGC 628 is an example of a galaxy that has a larger relative difference in mean pitch angles (see Figure \ref{fig2} and Table 2). Visually confirming the accuracy of our pitch angle measurements, can be done by performing an inverse fast Fourier transformation \citep{Davis:2012} (Figure \ref{fig3}). \textsc{Spirality} is a \textsc{MATLAB} code designed to measure pitch angles using a template fitting approach. Within the measurement annulus, \textsc{Spirality} uses a variable inner radius method and focuses on a global best-fit pitch. For six galaxies, pitch angles were adopted from a previous study by \cite{Pour-Imani:2016}, which had also used the same techniques to measure the pitch angles. The results were visually verified again by using a Python script (\textsc{spiral\_overlay.py}, hereafter \textsc{OL Script}) \footnote[4]{ Python OL Script: https://github.com/ebmonson/2DFFTUtils-Module }.

\subsection{Astrometric Accuracy}
\begin{figure}
	\includegraphics[width=8.6cm]{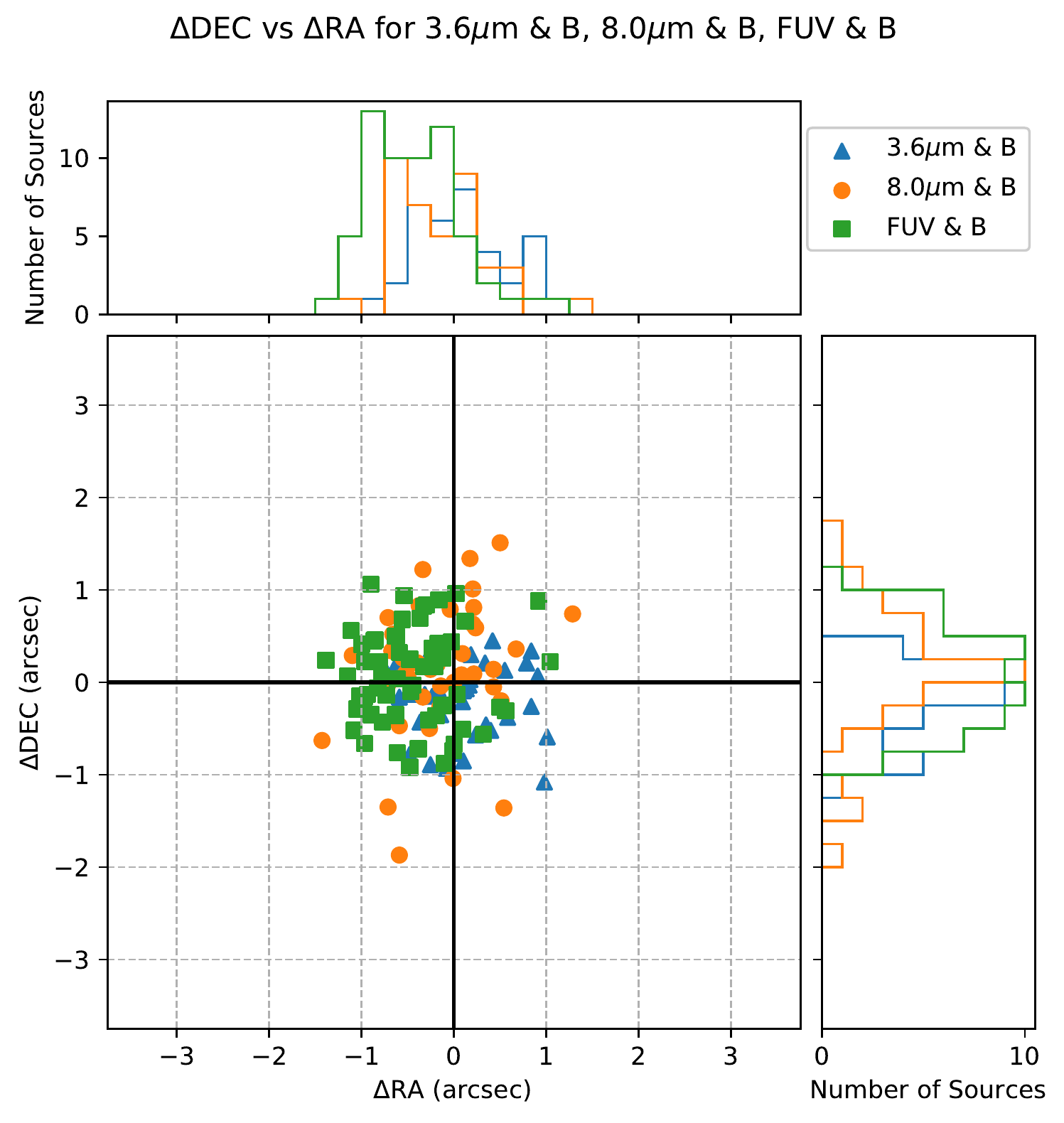}
	\caption{ Measured relative offsets of point sources in NGC 5194, for wavelengths 3.6$\mu$m, 8.0$\mu$m and $FUV$, compared against $B$-band. Considering the total data set, the mean of $\Delta RA$ is -0.2109 arcsec and the mean of $\Delta DEC$ is 0.0015 arcsec. }
	\label{fig4}
	
\end{figure}

Astrometric precision plays a vital role when overlapping images of different wavelengths and analyzing
their characteristics. Figure \ref{fig4} represents the offset distributions of point sources in NGC 5194, each wave
length compared against, $B$-band. As the initial process, point sources were identified in each image using
the DAOStarFinder class of the photutils package in \textsc{Astropy}. DAOStarFinder uses the DAOFIND \citep{Stetson:1987}
algorithm which searches images for local intensity maxima with peak amplitudes greater than a given
threshold value. After identifying the point sources, we searched for common points which were
detected in both wavelengths. Once we have a list of common point sources, we can find the
relative offsets in $\Delta RA$ and $\Delta DEC$. Considering the total data set, the mean of $\Delta RA$ is -0.2109 arcsec and the mean of $\Delta DEC$ is 0.0015 arcsec.

\begin{deluxetable*}{llccclllc}
\tablecolumns{7}
\tablecaption{Average Pitch Angle Measurements\label{Pitch Angles}}
\tablehead { \colhead{Galaxy Name} & \colhead{\textit{i $^{\circ}$}} & \colhead{\textit{PA $^{\circ}$ }} & \colhead{\textit{P $^{\circ}$} (IRAC 3.6) } & \colhead{\textit{P $^{\circ}$} ($B$-band)} & \colhead{ \textit{P $^{\circ}$} (IRAC 8.0)} & \colhead{  \textit{P $^{\circ}$} ($FUV$-band)} \\
\colhead{(1)} & \colhead{(2)} & \colhead{(3)} & \colhead{(4)} & \colhead{(5)} & \colhead{(6)} & \colhead{(7)} }

\startdata
NGC 0628 & 7 & 20 & $ 13.56 \pm 2.21 $ &  $ 13.94 \pm 0.64 $  & $ 12.83 \pm 0.66 $  & $ 12.81 \pm 0.65 $ \\
NGC 3031 &44 & 150 & $ 12.75 \pm 2.31 $ & $ 13.32 \pm 2.17 $ & $ 12.58 \pm 1.22 $  & $ 16.20 \pm 5.63 $ \\
NGC 2903 & 38.24 & 22.12 & $ 23.62 \pm 1.32 $ & $ 23.97 \pm 1.03 $ & $ 24.23 \pm 1.16  $ & $ 24.49 \pm 2.25 $ \\
NGC 4254 & 23.76 & 58.83 & $ 21.04 \pm 1.67 $  & $ 22.45 \pm 1.68 $ & $ 20.50 \pm 1.73 $   & $ 20.39 \pm 3.59 $ \\
NGC 5194 & 21.96 & 28.02 & $ 9.66 \pm 2.17 $  & $ 10.21 \pm 2.7 $  & $ 10.48 \pm 2.73 $  & $ 12.89 \pm 2.53 $ \\
NGC 5236 & $5.11$ & $-49.96$ & $ 10.56 \pm 1.78 $ & $ 10.63 \pm 1.27 $ & $ 11.71 \pm 1.95 $  & $ 12.92 \pm 1.10 $ \\
NGC 5457 & 7.64 & 4.01 & $ 22.77 \pm 1.24 $ & $ 23.33 \pm 1.28 $ & $ 23.85 \pm 1.65 $  & $ 24.13 \pm 2.94 $ \\
NGC 1566 & 12.12 & 178 & $ 19.95 \pm 1.84 $ & $ 18.23 \pm 1.74 $ & $ 22.37 \pm 1.54 $ & $ 18.90 \pm 1.99 $ \\
NGC 4321 & 25.9 & $-53.14$ & $ 16.07 \pm 2.42 $ & $ 16.05 \pm 2.75 $  & $ 14.73 \pm 1.68 $  & $ 14.40 \pm 1.58 $ \\
NGC 5033 & 43.71 & 4.26 & $ 7.09 \pm 0.46 $  & $ 10.46 \pm 2.66 $ & $ 13.91 \pm 4.42 $  &-&     \\
NGC 6946 & 9.2 & $-88.82$ & $ 22.02 \pm 1.02 $ & $ 24.10 \pm 1.8 $  & $ 24.63 \pm 2.01 $  &-& \\  
NGC 1042 & 43& 118 & $ 20.05 \pm 2.16 $ & $ 22.39 \pm 1.33 $  &  - & $  25.71 \pm 1.29 $\\
NGC 4579 & 46 & 75 & $ 20.35 \pm 1.24 $ & $ 18.18 \pm 1.16 $  &  - & $  24.8 \pm 1.27 $\\     
NGC 5701 & 54 & 174 & $ 9.38 \pm 1.03 $ & $ 8.08 \pm 1.57 $  &  - & $  10.52 \pm 1.35 $\\    
NGC 5850 & 24 & $-44$ & $ 8.46 \pm 0.14 $ & $ 7.39 \pm 1.47 $  & $ 13.22 \pm 0.30 $  &-&  \\    
NGC 3938 & 15 & 27 & $ 13.02 \pm 0.70 $ & $ 13.52 \pm 1.38 $  & $ 16.67 \pm 1.98 $  &-&  \\           
NGC 4136 & 2 & $-79$ & $ 10.07 \pm 1.06 $ & $ 12.06 \pm 2.16 $  &  - & $  15.68 \pm 3.16 $\\
NGC 7479 &18 & 37 & $ 14.14 \pm 1.21 $ & $ 15.74 \pm 1.49 $  &  - & $  16.47 \pm 3.37 $\\
NGC 7552 & 19 & $-76.98$ & $ 25.87 \pm 3.67 $ & $ 27.58 \pm 3.38 $  & $ 28.99 \pm 2.83 $  &-&  \\

\enddata

\tablecomments{Columns:
(1) Galaxy name;
(2) inclination angle in degrees
(3) Position angle in degrees
(4) pitch angle in degrees for Spitzer 3.6 $\mu$m;
(5) pitch angle in degrees for $B$-band;
(6) pitch angle in degrees for Spitzer 8.0 $\mu$m;
(7)  pitch angle in degrees for  $FUV$-band;
}
\end{deluxetable*}

\subsection{Tracing the Individual Images}
The overlaying technique consists of two main steps. The first step involves tracing the spiral arms on all four images (corresponding to the four wavebands). This process was done by two independent methods. One using the above mentioned \textsc{OL script} and the other utilizing, \textsc{Spirality; Spiral arm count}  \citep{Shield:2015} script. In order to incorporate the uncertainties of the pitch angles each image was traced twice, using the maximum and the minimum pitch angle uncertainties, giving rise to two image arrays. Since the tracing was done by using the maximum and the minimum image arrays, the uncertainties of the pitch angles play a vital role in this phase.  
\\
\subsubsection{Using the \textsc{Python OL script}}
\begin{figure}
\includegraphics[width=8.6cm]{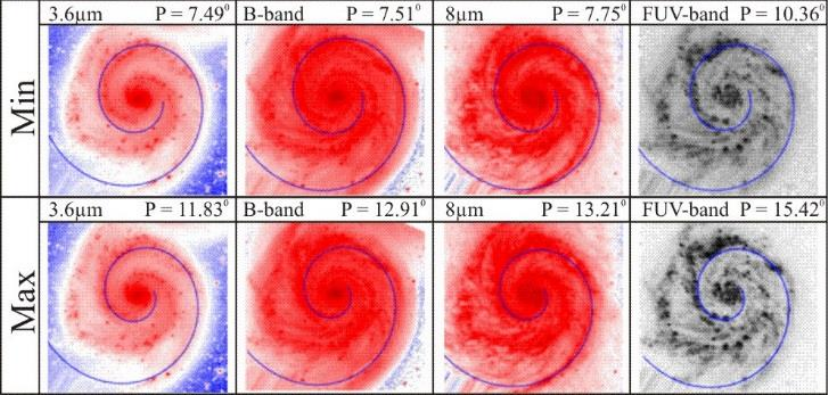}
\caption{Incorporating the uncertainties of the pitch angles, each image was traced twice, adding the positive and the negative uncertainties, giving rise to two image arrays corresponding to the maximum and minimum pitch angles. The \textsc{Python OL script} was used to create these images.}
\label{fig5}
\end{figure}

The \textsc{Python OL script} generates a graphical interface that enables us to load a FITS image, along with a synthetic logarithmic spiral placed on a foreground layer. Pitch angle, phase angle, the chirality (the direction in which galaxies appear to be) and the number of arms are among the parameters that are available for variation. Since the chirality, the number of arms, and pitch angles are known (through visual observations and using Table 2), the logarithmic spirals with the appropriate pitch angles can be precisely traced over the actual image by changing only the phase angle (see Figure \ref{fig5}). The accuracy of this process depends on the image clarity of the spiral arms and the contrast they show between the arm and inter-arm regions. As discussed, the extent to which the spiral arms are intrinsically logarithmic also plays an important role in the tracing process. Since the overlaying synthetic spiral traces are logarithmic, if the actual spiral structure of the galaxy has intrinsic deviations from being logarithmic, the overlaying would not be accurate.\\\\


\subsubsection{Using the \textsc{Spirality; spiral arm-count-script}}
\begin{figure}
\includegraphics[width=8.6cm]{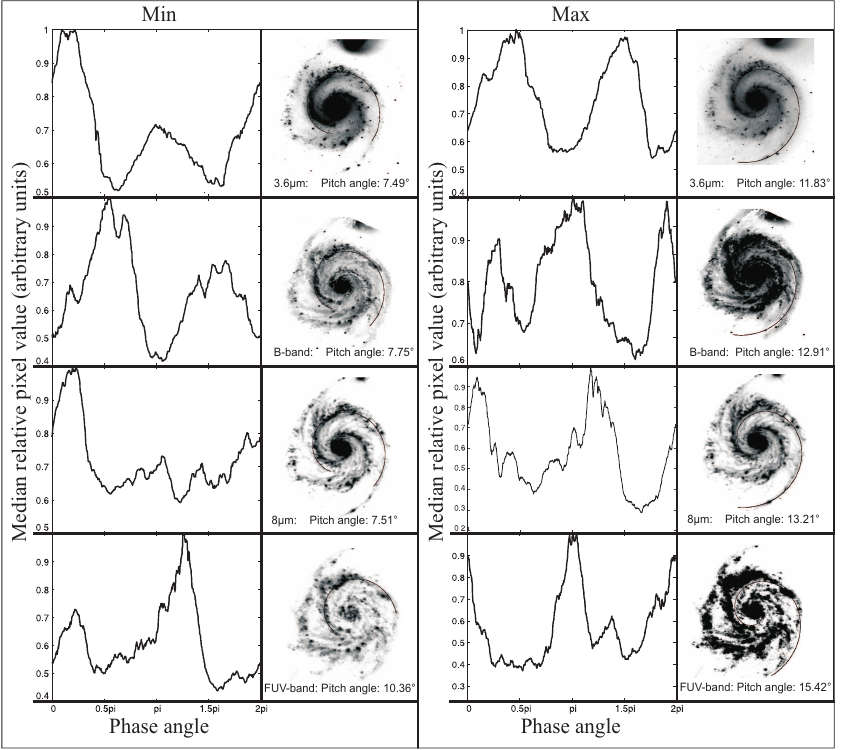}
\caption{Max and min image arrays produced using the \textsc{Spirality, spiral-arm-count} script. A graph of phase angle vs median relative pixel values is depicted, corresponding to each image. The locations of the maxima turning points correspond to bright regions of the spiral arms.                      }
\label{fig6}
\end{figure}

\textsc{Spirality, spiral-arm-count} \citep{Shield:2015} script is a MATLAB script that fits logarithmic spiral arms traced over images based on image pixel brightness. As the name suggests the \textsc{spiral-arm-count} script has the capability of measuring the number of arms in a given spiral galaxy image. It also produces a graph of phase angle vs median relative pixel values where the maxima turning points corresponds to the brightest regions of the spiral arms (see Figure \ref{fig6}).

\subsubsection{Trace Accuracy}
In order to quantify the accuracy of the tracing process, we defined the measurement ``Trace accuracy'', the ratio of the stable pitch angle region compared with the length of the spiral arm, expressed as a percentage.

$$Trace \ accuracy = \frac{(Stable \ pitch \ angle\ region)}{(Length \ of \ the \ spiral \ arm)  }\times 100 \%
$$

Higher trace accuracy simply implies a more reliable trace. Since the overlaying synthetic spiral traces are logarithmic, the extent to which the underlying galactic spiral arms are intrinsically logarithmic plays a vital role in having a higher trace accuracy. It is also important to note that well-defined spiral structures tend to have higher trace accuracies in comparison to flocculent galaxies. In our sample of 20 galaxies, four galaxies showed trace accuracies less than $40\%$, while NGC 2403 and NGC 1566 showed the lowest trace accuracies for all the bands.


\subsection{Overlaying}
Once we have all the individual images traced, the next step is overlaying the images and identifying the locations where the traced arms intercept each other. This process was done yet again by two independent methods. The first method involved the \textsc{Python OL script}, in which all the images were loaded as layers, saving one image at a time as a image. The second method used the RGB image creating capability of DS9 to construct the composite images. Individually traced images of different wavelengths were loaded as red, green and blue images in constructing the composite RGB image, giving preference to the images that gave the maximum outer extent of the R$_{CR}$ and the minimum inner extent of the R$_{CR}$. 

\section{Results}
\begin{figure}
\includegraphics[width=8.6cm]{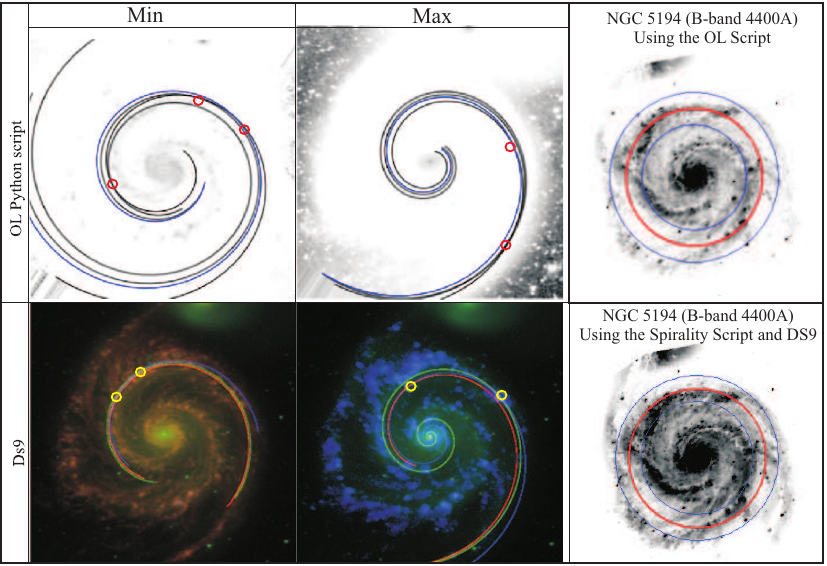}
\caption{The individually traced images are overlaid in the form of an RGB image using DS9. The locations in which the traced spiral arms intercept, correspond to the inner and the outer parts of the co-rotation radius.}
\label{fig7}
\end{figure}

\begin{figure}
\includegraphics[width=8.6cm]{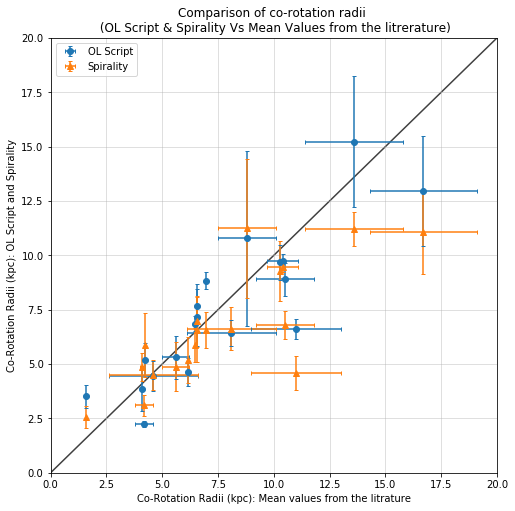}
\caption{A comparison of co-rotation radii, the blue data points indicates the results from the Overlaying \textsc{OL Script} method while the red data points indicates the results obtained from the \textsc{Spirality: Spiral arm count} method. The two results are compared against the mean values obtained from the literature (as summarized by  \cite{Buta:Zhang:2009} and \cite{Scarano:Lepine:2013}). }
\label{fig8}
\end{figure}

The median radii (depicted by the red circle in the
rightmost panels of Figure \ref{fig7}) were calculated by considering 
the maximum outer extent of the R$_{CR}$ and the
minimum inner extent of the R$_{CR}$ (depicted by the two
blue circles in the rightmost panels of Figure \ref{fig7}). The
R$_{CR}$ results that we obtained using the two independent
methods were compared against the mean values
compiled from the literature (see Figure \ref{fig8}). The mean
values obtained from the literature are as summarized by \cite{Scarano:Lepine:2013} and \cite{Buta:Zhang:2009}.  We can consider the mean
and the standard deviations of the normalized R$_{CR}$ values
as a measurement of the overall accuracy of our results.
Defining normalized R$_{CR}$ as

$$NR_{cr}(OL : Lit) = \frac{Rcr(Overlay)}{Rcr(Litrature)}
$$
\\
$$NR_{cr}(SP : Lit) = \frac{Rcr(Spirality)}{Rcr(Litrature)}
$$

For \textsc{OL Script} and \textsc{Spirality}, we found the mean normalized R$_{CR}$ values to be 0.997 and 0.966 respectively. The standard deviations were found to be 0.357 and 0.267 respectively.
Although few galaxies (e.g. NGC 5457, NGC 3031) showed significant deviations from the expected
values, most of our results were in agreement with the
compiled results. The final results are summarized in Table
3. The very fact that we were able to infer the R$_{CR}$
locations from this method serves as a compelling evidence
in favor of the implications of density wave theory.


\subsection{Time Lapse Calculations}
\begin{figure}
\includegraphics[width=8.7cm]{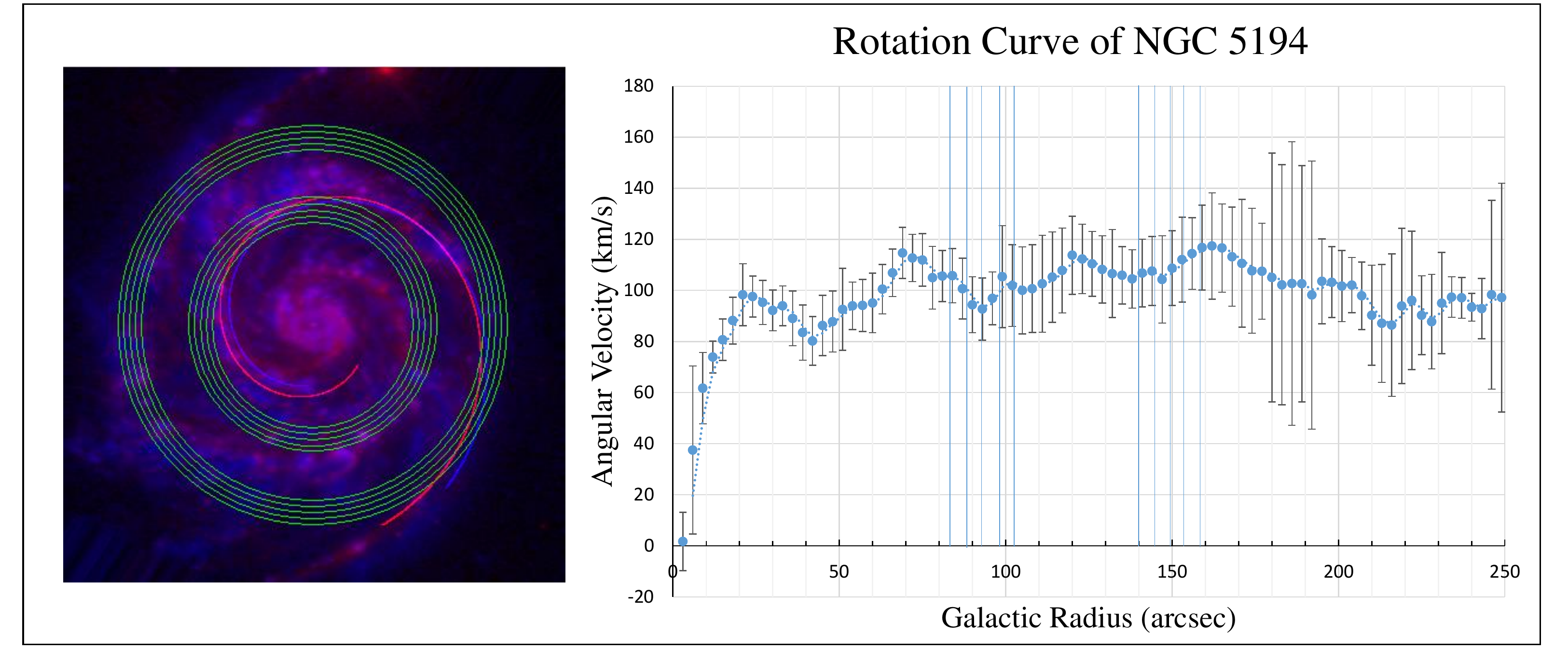}
\caption{(Left) The locations of the selected radii for the time lapse calculations on the galaxy NGC5194. 10 radii were selected both from inside and outside of the co-rotation radius. (Right) The rotation curve used, \cite{Daigle:2006}.}
\label{fig9}
\end{figure}

Once we had the co-rotation radius and the individual images with the spiral arms traced, we  used them to estimate the overall time lapse between the motion of  young arbitrary stars, born at the density wave, to move to their current locations. The time-lapse estimations were done using the 8.0$\mu$m and $B$-band traced images of six galaxies in our sample and using published rotation curves in the literature \citep{Daigle:2006}. Since the 8.0$\mu$m wavelength images trace emission from the dust and gas dominated regions, we may use it to portray the current location of the density wave. The $B$-band images can be used to locate the newly formed stars that are not obscured by dust. Assuming the azimuthal motion of the stars along the disk is circular in nature and assuming a constant pattern speed for the density wave, we analyzed the motion by selecting different radii, both from inside and outside the co-rotation radius (for a particular example consider NGC 5194, see Figure \ref{fig9} (left)). The angular velocity at the co-rotation radius was used as the global pattern speed of the density wave. The rotation
curves were used to identify the angular velocities
(see Figure \ref{fig9} (right)) and hence the relative velocities,
between the orbiting particles and the density wave, corresponding to each radii. Using the relative velocities,
we were able to estimate the overall mean time lapse for
this motion (see Table 4). It is important to note that we should exclude the region close to the co-rotation radius in this process. The overall mean time lapses
range from a minimum of 33.61 Myr to a maximum of
57.80 Myr. The average time lapse for this motion was
48.7 Myr. Depending on the mass of a star, an average
B-type star may have a life span of 10-100 million years,
hence our calculations are consistent with the current
established predictions.

It is important to mention that although we assumed the spiral structures to have a constant pattern speed, we did not discuss the validity of this assumption. \cite{Peterken:2019}, shows convincing evidence for a constant global pattern speed using $H \alpha$ regions to depict the ongoing star formations and considering the offsets with the young stars (less than 60Myr, traced by 4020$\AA$). They found that the pattern speed ($\Omega_P$) varies very little, with the radius and they further claim that the existence of the constant global pattern speed can be used as evidence in favor of the quasi-stationary density wave theory. 

\begin{deluxetable*}{llccclllc}

\tablecolumns{6}
\tablecaption{A comparison of our results against mean values obtained from literature \label{comparison}}
\tablehead{
 \colhead{Galaxy Name } & \colhead{Rcr: OL Script (kpc)} & \colhead{Rcr: SP Script (kpc)} & \colhead{ Rcr: Mean from Lit. (kpc)} & \colhead{$NR_{cr}(OL:Lit)$} & \colhead{$NR_{cr}(SP:Lit)$}} \\
 \colhead{(1)} & \colhead{(2)} & \colhead{(3)} & \colhead{(4)} & \colhead{(5)} & \colhead{(6)}

\startdata
NGC 0628 & $ 4.45 \pm 0.68 $  & $ 4.50 \pm 0.68 $   & $ 4.6 \pm 2 $ & $0.97$ & $ 0.76$ \\
NGC 3031 & $ 6.61 \pm 0.46 $  & $ 4.59 \pm 0.80 $   & $ 11 \pm 2 $ & $0.60$ & $0.54$ \\
NGC 2903 & $ 3.87 \pm 1.02 $  & $ 4.86 \pm 0.64 $   & $ 4.1 \pm 0.1 $ & $0.94$ & $1.19$ \\
NGC 4254 & $ 15.23 \pm 3.00 $ & $ 11.22 \pm 0.78 $  & $ 13.6 \pm 2.2 $ & $1.12$ & $1.17$ \\
NGC 5194 & $ 5.31 \pm 0.98 $  & $ 4.88 \pm 1.12 $   & $ 5.6 \pm 0.6 $ & $0.95$ & $0.87$ \\
NGC 5236 & $ 6.44 \pm 0.61 $  & $ 6.63 \pm 1.00 $   & $ 8.1 \pm 2 $ & $0.79$ & $0.82$ \\
NGC 5457 & $ 12.94 \pm 2.53 $ & $ 11.06 \pm 1.9 $   & $ 16.7 \pm 2.4 $ & $0.77$ & $0.66$ \\
NGC 1566 & $ 10.78 \pm 4.01 $ & $ 11.24\pm3.21 $    & $ 8.8 \pm 1.3 $ & $1.22$ & $1.28$ \\
NGC 4321 & $ 8.90 \pm 0.75 $  & $ 6.81 \pm 0.64 $  & $ 10.5 \pm 1.3 $ & $0.85$ & $1.14$ \\
NGC 5033 & $ 9.75 \pm 0.33 $  & $ 9.48  \pm  0.24 $ & $ 10.4 \pm 0.7 $ & $0.94$ & $0.91$ \\
NGC 6946 & $ 2.26 \pm 0.14 $  & $ 3.11 \pm 0.48 $   & $ 4.2 \pm 0.4 $ & $0.54$ & $0.74$ \\
NGC 1042 & $ 4.64 \pm 0.63 $  & $ 5.20 \pm 1.06 $   & $ 6.14 * $ & $0.76$ & $0.85$ \\  
NGC 4579 & $ 7.15 \pm 1.31 $  & $ 6.58 \pm 1.49 $   & $ 6.58 * $ & $1.09$ & $1.00$ \\    
NGC 5701 & $ 6.86 \pm 0.35 $  & $ 5.88 \pm 0.77 $   & $ 6.48 * $ & $1.06$ & $0.91$ \\   
NGC 5850 & $ 8.84 \pm 0.38 $  & $ 6.57 \pm 0.84 $   & $ 6.96 * $ & $1.27$ & $0.94$ \\
NGC 3938 & $ 7.68 \pm 1.00 $  & $ 6.97 \pm 1.17 $   & $ 6.55 * $ & $1.17$ & $1.06$ \\ 
NGC 4136 & $ 3.51 \pm 0.51 $  & $ 2.55 \pm 0.50 $   & $ 1.60 * $ & $2.19$ & $1.59$ \\  
NGC 7479 & $ 9.68 \pm 0.80 $  & $ 9.27 \pm 1.37 $   & $ 10.29 * $ & $0.94$ & $0.90$ \\ 
NGC 7552 & $ 5.17 \pm 0.78 $  & $ 5.86 \pm 1.47 $   & $ 4.23 * $ & $1.22$ & $1.39$ \\    
      
\enddata

\tablecomments{Columns:
(1) Galaxy name;
(2) Results from the Overlaying \textsc{OL Script} method;
(3) Results from the \textsc{Spirality: Spiral arm count} method;
(4) Mean values obtained from the literature (as summarized by \citep{Scarano:Lepine:2013} and * \citep{Buta:Zhang:2009});
(5) Normalized $R_{cr}$ for OL Script vs Mean from Lit  $NR_{cr}(OL:Lit)$ ;
(6) Normalized $R_{cr}$ for SP Script vs Mean from Lit  $NR_{cr}(SP:Lit)$. 
}

\end{deluxetable*}

\begin{deluxetable*}{llccclllc}
\tablecolumns{5}
\tablecaption{Time lapse calculations\label{Time lapse calculations}}
\tablehead{
\colhead{Galaxy Name } &  \colhead{Rcr: Spirality Script (kpc)} & \colhead{ $\Omega$$_{p}$  (km/s/kpc)} & \colhead{Mean Relative Velocity (pc/Myrs)} & \colhead{Mean Time Lapse (Myrs)} \\
\colhead{(1)} & \colhead{(2)} & \colhead{(3)} & \colhead{(4)} & \colhead{(5)}}

\startdata
NGC 0628 & $ 4.50 \pm 0.68 $ & $ 33.51 \pm 10.65 $ & $ 44.24 \pm 15.05 $ & $ 33.61 \pm 3.84 $ \\
NGC 3031 & $ 4.59 \pm 0.80 $ & $ 40.23 \pm 4.87 $ & $ 28.24 \pm 3.63 $  & $ 48.62 \pm 2.73 $ \\
NGC 5194 & $ 4.88 \pm 1.12 $ & $ 21.56 \pm 6.13 $ & $26.84 \pm 8.35 $  & $ 54.62 \pm 8.75$\\
NGC 4321 & $ 6.81 \pm 0.64 $  & $ 26.21 \pm 4.62 $ & $ 31.42 \pm 5.73 $  & $ 43.86 \pm 2.69 $\\
NGC 6946 & $ 3.11 \pm 0.48 $ & $ 51.78 \pm 10.00 $ & $ 13.63 \pm 2.95$ & $ 57.80 \pm 3.79 $\\

\enddata

\tablecomments{Columns:
(1) Galaxy name;
(2) Rcr: Results from the \textsc{Spirality: Spiral arm count} method;
(3) Patern Speed $\Omega$$_{p}$ ;
(4) Mean relative velocity of a young arbitrary star that was born at the density wave (peak intensity isophotes of the 8$\mu$m image) to the current location (peak intensity isophotes of the $B$-band image);
(5) The overall mean time lapses for the motion of a young arbitrary star that was born at the density wave (peak intensity isophotes of the 8$\mu$m image) to the current location (peak intensity isophotes of the $B$-band image).
}
\end{deluxetable*}

\subsection{Comparison with Results from Other Research}
As we have seen, using spiral arm pitch angle measurements at different wavelengths as a method to find the co-rotation radius agrees quite well with established techniques. Naturally, this tends to confirm the existence of a color gradient in spiral arms in disk galaxies. It lends confidence in our ability to accurately distinguish between the upstream and the downstream spiral arms. Nevertheless, there are skeptics who doubt the existence of such a gradient. Some researchers argue that there is no consistent color gradient even within local stretches of spiral arms in a given galaxy. By contrast, several studies, which use pitch angle measurements to try to study the entire length of the spiral arm, agree that there is evidence for a consistent color gradient. Recently a particularly large study by \cite{Yu_2018} added to the growing body of evidence on this point. They find a consistent color gradient and present a model which illustrates how a red and a blue spiral arm (called by them respectively, the potential or P arm and the star-formation or SF arm) will cross each other at the co-rotation radius. In their model the
downstream arm contains stars which have moved from the upstream star-forming region over a timescale of 10 Myr or so.

In one respect, \cite{Yu_2018} disagree with the earlier results reported in \cite{Pour-Imani:2016}/\cite{Rayan:2019}, find that the infrared pitch angle is generally looser (not tighter, as we find in this paper, in agreement with \cite{Rayan:2019}) than the pitch angle as measured in other wavelengths, such as $B$-band. This is consistent with the red spiral arm being generated by old disk stars crowded closer together by the peak of the density wave\textquoteright s spiral potential. \cite{Yu_2018} calls this the color gradient from red to blue and we in contrast see what they call an age gradient, of young stars seen downstream from the star-formation region, from blue to red. Yu and Ho contend that the age gradient is not really visible and believe that our measurements of infra-red pitch angles are marred by errors. In one particular case they are undoubtedly right. NGC 1566 is a galaxy found in both their sample and the one in \cite{Pour-Imani:2016}, for which the pitch angle measurements are in sharp disagreement. \cite{Yu_2018} argue that this is because of a failure to properly choose consistent galaxy parameters, such as galaxy center, ellipticity, position angle, and radius range throughout the different wavelengths that are used, and they are right. We discovered that for six of the galaxies in that sample, the inclinations that we used were not consistent with the ones used by \cite{Yu_2018}. For one of these galaxies (NGC 1566) redoing the measurements more carefully with consistent galaxy parameters results in a change of values which agrees better with Yu and Ho\textquoteright s results, and with their findings on the color gradient. However, we should also point out, as \cite{Yu_2018} did not, that NGC 1566 is not the only galaxy common to the samples of their paper and \cite{Pour-Imani:2016}. NGC 1097 is also, and in that case our figures agree quite well and the color gradient observed is in agreement with the trend we see (though not strongly).

\begin{figure}
	\includegraphics[width=8.7cm]{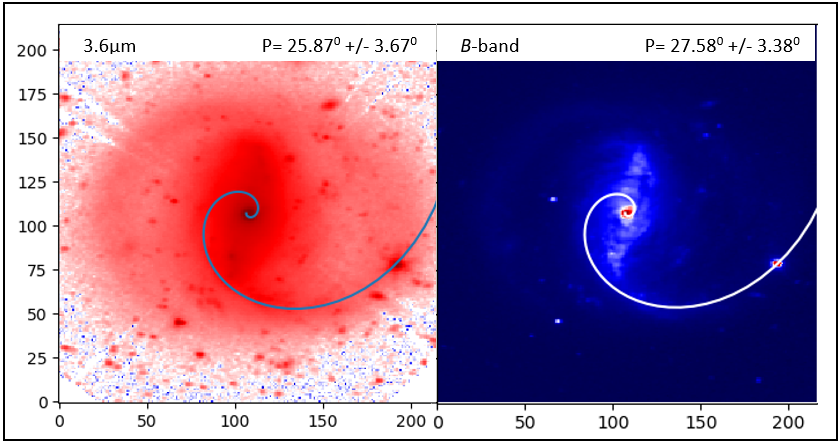}
	\caption{(left) 3.6$\mu$m image of NGC 7552 along with a synthetic logarithmic spiral of pitch angle $25.87 \pm 3.67^{\circ}$ traced over the brightest arm.
	(right) $B$-band image of NGC 7552 along with a synthetic logarithmic spiral of pitch angle $27.58 \pm 3.38^{\circ}$ traced over the brightest arm .}
	\label{fig10}
\end{figure}

Another galaxy which is common to \cite{Yu_2018} and our sample is NGC 7552. In this case there is also a considerable disagreement for the 3.6$\mu$m and $B$-band measurements, since the values given in \cite{Yu_2018} are $9.7 \pm 1.1^{\circ}$ and $9.7 \pm 0.9^{\circ}$ respectively while our values are $25.87 \pm 3.67^{\circ}$ and $27.58 \pm 3.38^{\circ}$ (Figure \ref{fig10}). However \cite{Yu_2018} gives only measurements for their 1DFFT method. Apparently their 2DFFT algorithm did not return a value for this galaxy, which might be because it has a long bar. Both of our methods (\textsc{2DFFT} and \textsc{Spirality}) agree in this case. In order to check further we made use of an entirely independent code, SpArcFiRe, which returned a value of $27.03 \pm 3.85^{\circ}$ for 3.6$\mu$m, compatible with our results, but not with the values given by \cite{Yu_2018}. Although it is not always 
trivial to compare results between these papers, and although there is a consistent disagreement over pitch angle measured in the near infrared (3.6$\mu$m), the broader picture of distinct upstream and downstream arms emerging from both studies is well confirmed by the work in this paper, in which this result is used successfully to locate the co-rotation radius of a significant sample of disk galaxies. Nevertheless inconsistencies such as the disagreement on the location of the 3.6$\mu$m infrared arm will provide some comfort for those who favor the transient picture of spiral arms.   

Obviously, further work (and more careful work!) is needed to decide this issue. But notice that the disagreement over the
pitch angle of the infrared spiral arm is not particularly
relevant to the work discussed in this paper.  We qualitatively (and wholeheartedly) agree with the model presented, for instance, in Fig. 6 of \cite{Yu_2018}, except that we would use star-formation tracers such as FUV or 8$\mu$m  to identify the upstream arm (their P arm) and $B$-band images to find the downstream arm (their SF arm). The spirit of this paper, therefore, is actually to see if the measured spiral arms actually perform as one would expect given a model similar in spirit to the one discussed in section 4.3 of \cite{Yu_2018}. Our answer is an emphatic yes. Invariably tracing of the two spiral arms measured by us does correctly identify the position of the co-rotation radius. We find the elapsed time between the two arm positions to be something in the range of 50 Myr, not too far from the 10 Myr assumed in model presented in \cite{Yu_2018}. Therefore, in spite of certain disagreements in detail, we find that the evidence is quite strong for the idea that the color gradient predicted by density wave theory is real. Of course, continued disagreement over the detail leaves an opening for those who argue that there is no consistent age/color gradient. This issue is important because it has been proposed, \cite{Yuan:1981} that a consistent age/color gradient supports the theory that spiral arms are long-standing, while no consistent color gradient supports the theory that spiral arms are transient.

Recent work looking for observational evidence on this subject has been 
divided. \cite{Foyle:2011} take the 
view that there is no evidence for the
sort of color-gradient structure that we see and conclude that spiral structure is 
transient (their work differs from ours in examining local structure within each arm,
whereas we look at the full extent of the spiral arm). By contrast, \cite{MartinezGarcia:GonzalezLopezlira:2015}  reach the opposite conclusion, using similar
methods. Our work, though very different in approach, seems to be more reconcilable with the Martinez-Garcia and Gonzalez-Lopezlira conclusion.

\section{Summary and Conclusion }
We have confirmed the validity of density wave theory in a very direct way by using differently colored spiral arms
to find the co-rotation radius and showing that it agrees well with other methods commonly used. This shows the
existence of a color gradient demanded by the theory and shows that the spiral arm patterns themselves can be used
as a research tool. It therefore is implicit that the arms have a certain reality which is suggested by their success in
measurement.

We have compared our results with several methods used to identify the location of the co-rotation radius in a galaxy,
with considerable success. It is true that errors in the pitch angle measurements translate to considerable errors in
the measurement of the co-rotation radius. We do not propose that our method should supplant other methods of
finding the co-rotation radius. But it is clear that our method is usually accurate and always qualitatively plausible.
Even when it is not perfectly accurate, it is not too far off. The argument is that the success in finding the co-rotation radius, that these arms are genuine features created by
the underlying density wave. 

A major point of debate about the density-wave theory concerns how long
the spiral patterns generated by density waves last. 
The original Lin-Shu
theory proposed that the spiral arms were semi-permanent standing-wave
patterns, created by resonant modes 
reflecting back and forth between the
inner and outer Lindblad resonances of the disk. Most experts no longer think
this is likely. The system seems too dissipative for standing waves to persist
indefinitely. Certainly this is what numerical 
simulations seem to suggest.
Indeed it is unclear how such disk-spanning patterns can be excited at all,
though the favored 
mechanisms are external harassment (by another galaxy
passing nearby and stimulating a disturbance) or swing amplification (by
which small internal perturbations might be amplified) \citep{Toomre:1969}. Since it is empirically
apparent that the patterns do form, the key 
question is how long they last.

Some experts, such as Sellwood and Carlberg \citep{Sellwood:Carlberg:2014}, argue that the patterns can be
quite long-lasting, perhaps for as many as ten galactic rotations. But there
is quite a common belief that they may be quite transient, perhaps lasting
only one or two rotations of the disk. 
Our work suggests that it takes on the order of 50 Myr for the spiral pattern
defined by the $B$-band to form. When that stellar arm has formed, the original star-forming
region can still be seen. This suggests that the spiral arm is a minimum of
50 Myr old. Obviously this is less then a galactic rotation (the Milky Way
takes about 250 Myr to rotate once). However, for most of our galaxies
we do see a double pattern of this type, which suggests that it is rare to
catch a galaxy as the density wave pattern is breaking up. Though we cannot
be sure, this does suggest that the typical age of a spiral pattern is several
times 50 Myr. This does at least place some lower limit on how transient
these patterns can be. It is hard to see how they can generally persist only
for one rotation, and even two might be a reasonable lower limit.

Finally, it should be noted that the overall mean time lapses for the motion of a young star from the star-forming
region to the $B$-band is consistent with our evolutionary understanding of this process. We would expect the
time involved to be longer than 10 million years, so that the shortest-lived O type stars would still be seen in the
immediate vicinity of the star-forming region, since they live only a few million years. This is consistent with the fact
that our far UV images return pitch angles consistent with the far IR pitch angles ( see \cite{Rayan:2019}, for a more detailed discussion). At the same time, we expect that the time elapsed will be less than 100 million years, so that many bright stars will survive to be observable when it is viewed in the $B$-band. All of the values
we have obtained for the time elapsed to move between the star-formation arm and the $B$-band lie between
these two values, in the tens of million years, with an average close to 50 Myr (see Table 4). We believe that
the overall interpretation we give of our results is consistent with relatively long-lasting density waves.
\\

\acknowledgments
The authors gratefully acknowledge Bret Lehmer, Rafael Eufrasio, Marc Seigar and Joel Berrier for giving suggestions and contributing to this paper in numerous ways. We would like to thank Iv\^{a}nio Puerari for writing the original 2DFFT code on which the later versions of 2DFFT was based. We would also like to thank all the members of the Arkansas Galaxy Evolution Survey (AGES) team for their continuous support. This research has made use of the NASA/IPAC Extragalactic Database (NED) and the NASA\textquoteright s Astrophysics Data System.

\end{document}